\documentstyle[multicol,aps]{revtex}
\input{psfig}
\begin{document}
\newcommand{\la} {\langle}
\newcommand{\ra} {\rangle}
\newcommand{\ep} {\varepsilon}
\newcommand{\bu} {\bf u}
\newcommand{\de} {\delta}
\pagestyle{myheadings}
\markright{Submitted to PRL}
\draft
\preprint{Submitted to Phys. Rev. Lett.}
\title{A Refined Similarity Hypothesis for
Transverse Structure Functions}
\author{Shiyi Chen$^{1,2}$, Katepalli R. Sreenivasan$^{3}$, 
Mark Nelkin$^{4}$ and Nianzheng Cao$^{1}$}
\address{${}^{1}$IBM Research Division, T. J. Watson Research Center,
P.O. Box 218, Yorktown Heights, NY 10598\\
${}^{2}$Theoretical Division and Center for Nonlinear Studies,
Los Alamos National Laboratory, Los Alamos, NM 87545\\
${}^{3}$Mason Laboratory,  Yale University, New Haven, CT  
06520-8286\\
${}^{4}$Physics Department, New York University, New York, NY 10003\\
and Levich Institute, CCNY, New York, NY 10031 USA}

\maketitle

\begin{abstract}

We argue on the basis of empirical data that Kolmogorov's refined  
similarity hypothesis (RSH) needs to be modified for transverse  
velocity increments, and propose an alternative. In this new  
form, transverse velocity increments bear the same relation to  
locally averaged enstrophy (squared vorticity) as longitudinal  
velocity increments bear in RSH to locally averaged dissipation. We  
support this hypothesis by analyzing high-resolution numerical  
simulation data for isotropic turbulence. RSH and its proposed  
modification for transverse velocity increments (RSHT) appear to  
represent two independent scaling groups.

\end{abstract}

\begin{multicols}{2}
\narrowtext

In all small-scale turbulence research driven by expectations of  
universality, one deals with so-called velocity increments, which are  
differences of velocities between two spatial positions separated by  
a fixed distance. Of special interest are velocity increments when  
the separation distances belong to the inertial range, that is,  
length scales which are small compared to a typical large-scale  
motion but large compared to a typical viscous cut-off scale. It is  
fair to say that much of the phenomenological work on this subject,  
summarized in Refs. \cite{monin,frisch,sreeni}, is based to some  
degree or another on two sets of similarity hypotheses proposed by  
Kolmogorov, abbreviated as K41 \cite{k41} and K62 \cite{k62}. The  
latter, also called the refined similarity hypotheses (RSH), have  
been verified both in real experiments and numerical simulations of  
turbulence---at least to an extent that certifies them as reasonable.  
The verification has focused, largely for historical reasons of  
experimental convenience, on longitudinal velocity increments, that  
is, velocity increments for which the separation distance is aligned  
with the velocity component considered. The theme of this Letter is  
that RSH is inadequate for transverse  
velocity increments, for which the separation distance is transverse  
to the direction of the velocity component considered, and that a  
non-trivial modification (RSHT) is necessary to account for  
experimental facts. We conclude that RSH and RSHT form two  
independent scaling groups in small-scale turbulence.

We need a few definitions before proceeding further. The rate of  
energy dissipation is given by $\ep = \frac{\nu}{2} (\partial  
u_i/\partial x_j +  
\partial u_j/\partial x_i)^2$, where $\nu$ is the fluid viscosity,  
and its local average is defined as  
$\ep_r({\bf x},t) = \int_{V_r} \ep~dV$, where the integration volume  
$V(r)$, with the characteristic linear dimension $r$, is centered at  
the spatial position ${\bf x}$. The longitudinal velocity increment is defined as  
$\delta u_r \equiv u(x+r) - u(x)$, where the velocity component $u$  
and the separation distance $r$ are both in the same direction, say $x$. The  
form of RSH that has been verified extensively  
\cite{sreeni1,pras,chen1} is given by the relation
\begin{equation}
\de u_r = \beta_1 (r\ep_r)^{1/3},
\label{rsh1}
\end{equation}
where $\beta_1$ is a stochastic variable independent of $r$ and  
$\ep_r$. Assume now that the $p$-th order longitudinal structure  
function $S_p^{L}(r) \equiv \la (\de u_r)^p \ra \sim r^{\zeta_p^{L}}$  
and $\la \ep_r^p \ra \sim r^{\tau_p}$. Here the angular brackets  
denote suitable averages. It then follows from Eq.  
(\ref{rsh1}) that
\begin{equation}
\zeta_p^{L} = \frac{p}{3} + \tau_{p/3}.
\label{rsh1-e}
\end{equation}
This equation connects the scaling exponents of the longitudinal  
structure functions with those of the locally averaged dissipation  
function. If a further assumption about the statistics of $\ep_r$ can  
be made, such as log-normal \cite{k62}, multifractal \cite{mene}, or  
log-Poisson \cite{she}, $\tau_p$ can be obtained analytically via  
Eq. (\ref{rsh1-e}). 

The thinking in K62 is that Eq. (1) holds equally well if the  
longitudinal velocity increment is replaced by a transverse velocity  
increment, namely $\de v_r = v(x+r) - v(x)$, where the separation  
distance $r$ is transverse to the velocity component $v$. If true,  
this would imply that the scaling exponents $\zeta_p^{T}$  
and $\zeta_p^{L}$, for longitudinal and transverse structure  
functions, respectively, are equal. A kinematic constraint from  
isotropy \cite{monin,nelkin} assures us that $\zeta_2^{T} = \zeta_2^{L}$. Two  
sets of measurements \cite{camussi} appear to suggest (or imply) 
that the equality holds,  
within experimental uncertainty, for larger $p$ as well. On the other  
hand, more recent numerical \cite{boratav,grossmann} and experimental 
work \cite{benzi} has revealed that transverse velocity increments are  
more intermittent than longitudinal ones, and that $\zeta_p^T$ are  
measurably smaller $\zeta_p^L$, at least for $p > 4$. These results  
imply that the RSH cannot  
be equally true for both longitudinal and transverse velocity  
increments. With this in mind, we first obtain the  
scaling exponents of longitudinal and transverse structure functions.  
We employ the numerical data from simulations of isotropic turbulence,  
carried out using $512^3$ mesh points in a periodic box; a  
statistically steady state was obtained by forcing low Fourier modes.  
For details see the paper cited first in \cite{chen1}. The Taylor  
microscale Reynolds number $R_{\lambda} = 216$, which is close to  
the limit of the current computational capability. Averages over 10  
large-eddy turnover times were used in the data analysis. 

Figure 1 shows the flatnesses of longitudinal and transverse velocity  
increments as functions of $r$. For all values of $r$ except perhaps  
those comparable to the box size, the velocity increment along the  
transverse direction has a larger flatness, suggesting that the  
transverse velocity increment is more intermittent \cite{grossmann}.  
This shows that significantly more averaging is required for  
transverse quantities than for longitudinal quantities.

\bigskip
\psfig{file=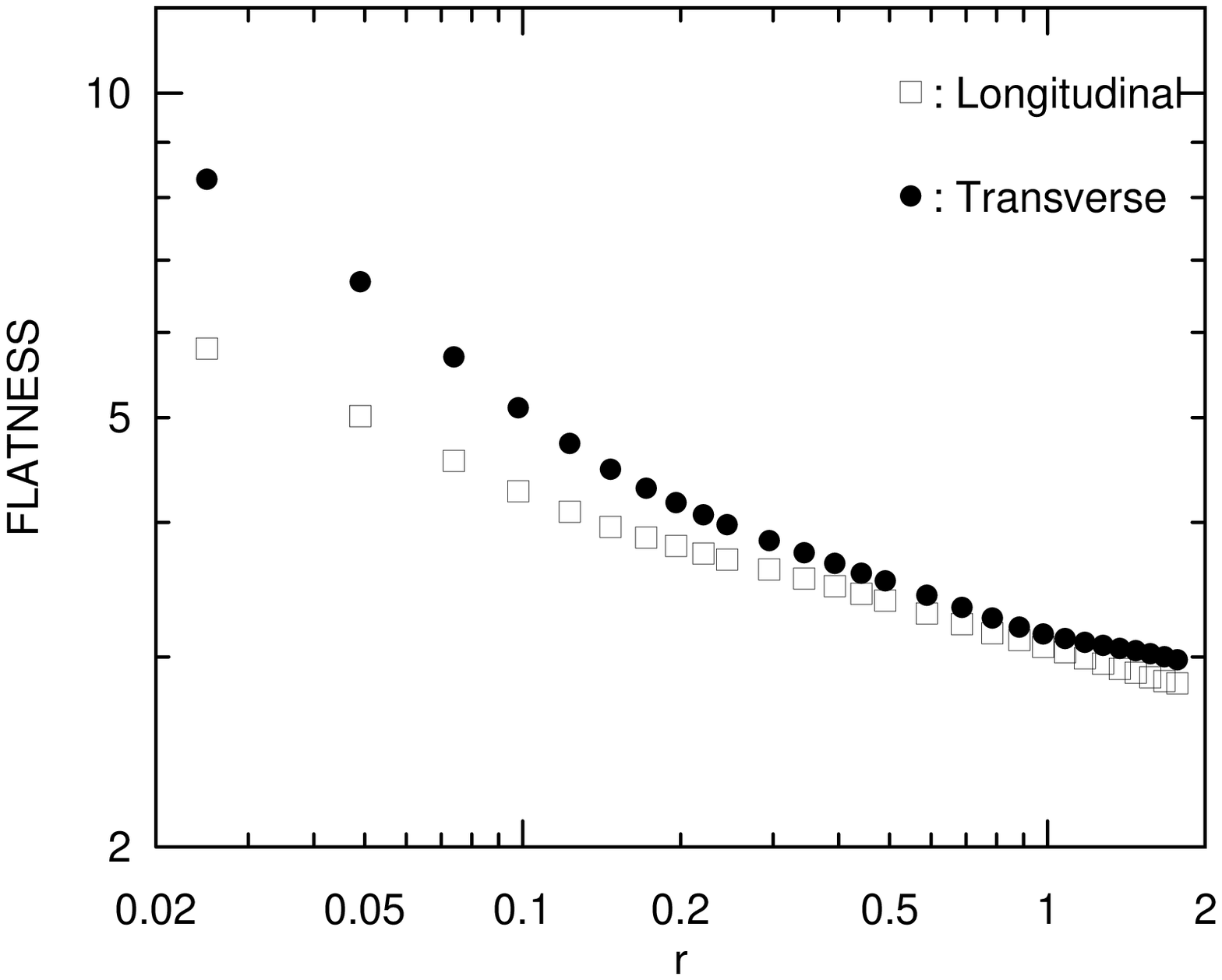,width=200pt}
\noindent
{\small FIG.~1. Flatnesses of velocity increments,
$S_4^{L}(r)/S_2^{L}(r)^2$ and $S_4^{T}(r)/S_2^{T}(r)^2$, as
functions of $r$.}
\bigskip

In Fig.~2, the transverse structure functions $S^T_p(r)$ are plotted  
as functions of $r$ for $p = 2, 4, 6, 8, 10$. An inertial-range 
power-law scaling can be identified for $0.2 \leq r \leq 0.6$  
(the whole box size being $2\pi$). This is also the scaling range  
determined \cite{sreeni2} from Kolmogorov's 4/5 law for the  
third-order structure function \cite{kolm}. 

\bigskip
\psfig{file=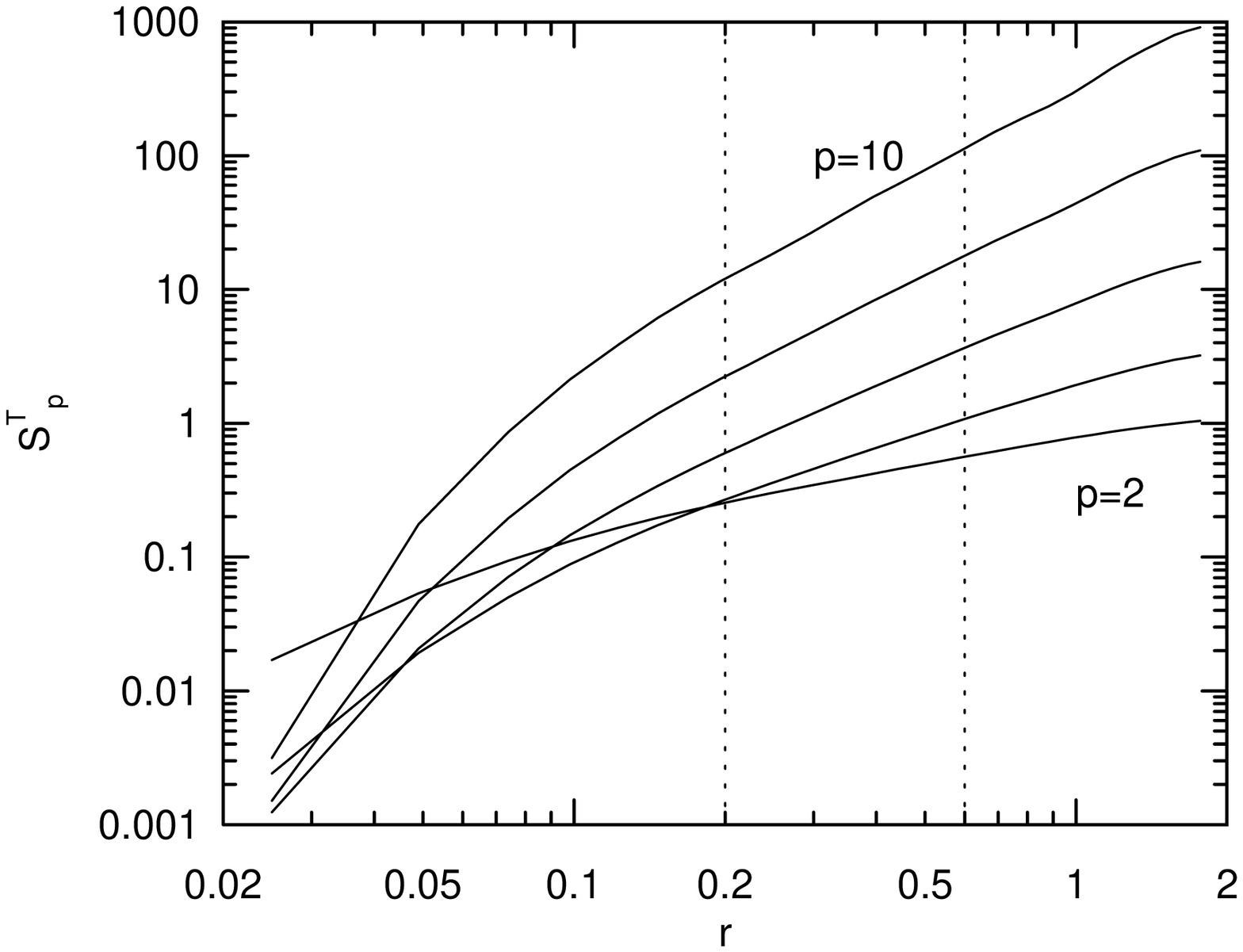,width=200pt}
\noindent
{\small FIG.~2.
The transverse structure functions $S_p^{T}(r)$ as functions of $r$
for $p = 2, 4, \cdot\cdot\cdot, 10$. }
\bigskip

The exponents $\zeta_p^T$, determined by least-square fits within the  
inertial range just mentioned, are plotted in Fig.~3. The use of   
Extended Self-Similarity (ESS) technique \cite{ess} yields
very similar results \cite{ess1}. The results for $\zeta_p^L$,  
obtained earlier\cite{cao}, have also been plotted for comparison. These  
results demonstrate that while both exponents show considerable  
anomaly due to intermittency effects (as evidenced by the deviation  
from the dotted line given by K41), the transverse exponents are  
systematically smaller than the longitudinal ones for $p >3$. Typical  
scaling exponents for $S_p^{T}$ are: $\zeta^T_2 = 0.71 \pm 0.04$,  
$\zeta^T_4 = 1.25 \pm 0.067$, $\zeta^T_6 = 1.63 \pm 0.079$ and  
$\zeta^T_8 = 1.87 \pm 0.078$. While the longitudinal exponents agree  
quite well with existing scaling models  
\cite{mene,she,van}, the difference between the transverse exponents  
and the models increases with $p$. For brevity, we show only a  
comparison with the log-Poisson model \cite{she}, which is typical. 

\bigskip
\psfig{file=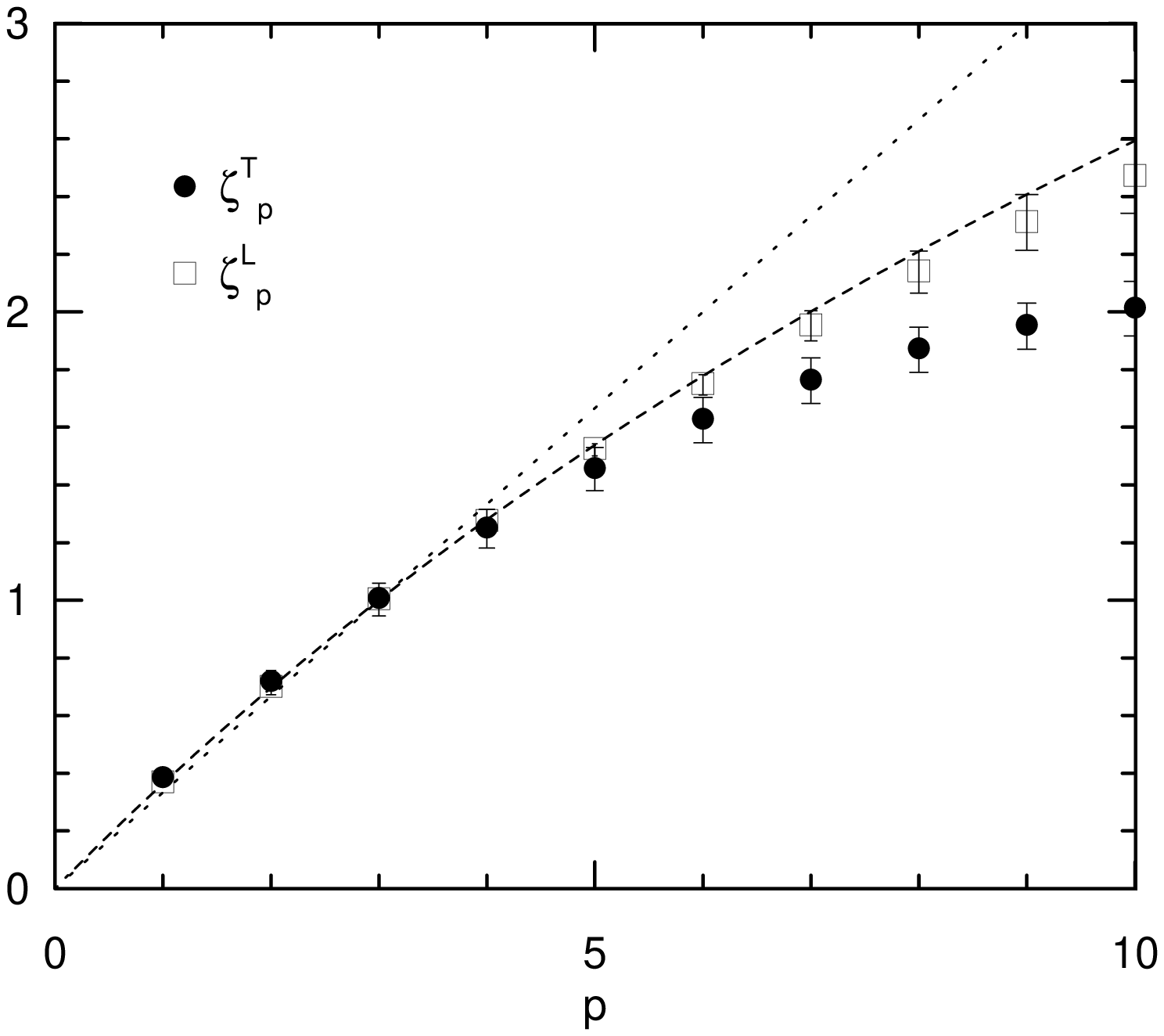,width=200pt}
\noindent
{\small FIG.~3. Numerical results for the transverse scaling
exponents, $\zeta_p^{T}$ and 
$\zeta_p^{L}$, as functions of $p$. The dotted line is for the
normal scaling relation (K41) and the dashed line is for the
log-Poisson model \cite{she}.}
\bigskip

If the scaling exponents for longitudinal and transverse structure  
functions are different, it is clear that Eq. (1), now known to be  
valid for the former, cannot be valid for the latter; some  
modification will therefore be required. One possibility suggests  
itself from kinematic considerations. Recall from the second  
reference of \cite{sreeni1} that RSH may be expected because the  
longitudinal velocity increment $\delta u_r$ is an integral over the  
separation distance $r$ of the velocity derivative $\partial  
u/\partial x$, whose square is one of the components of the energy  
dissipation. In a similar way, the transverse structure function  
$\delta v_r$ is the integral of $\partial v/\partial x$, which is one  
of the two terms of the vorticity component $\omega_z = \partial  
v/\partial x - \partial u/\partial y$. It is reasonable to guess,  
then, that a suitable modification of Eq. (1) would be
\begin{equation}
\de v_r = \beta_2 (r{\Omega}_r)^{1/3},
\label{rsh2}
\end{equation}
where $\Omega = \nu \omega^2$, $\omega = \nabla \times {\bf u}$ is  
the vorticity, and $\Omega_r$ is the local average of $\Omega$ 
in the same way as $\ep_r$ is the local average of $\ep$ in  
Eq. (\ref{rsh1}). A necessary condition for the above equation to be  
plausible is that $\ep_r$ should scale differently from $\Omega_r$  
and that the difference must be compatible in sign with that of the  
difference between $\zeta_p^L$ and $\zeta_p^T$. That this is indeed  
so was demonstrated recently \cite{vor}; that is, if $\la \Omega_r^p  
\ra \sim r^{o_p}$, $o_p < \tau_p$.  Thus, Eq. (3)  
is {\em a priori} worth exploring as a candidate for the refined  
similarity hypothesis for transverse velocity increments (RSHT). 

We now test the validity of Eq. (3). In particular, we test whether  
the relation
\begin{equation}
\zeta_p^{T} = \frac{p}{3} + o_{p/3},
\label{rsh2-e}
\end {equation}
which follows from Eq. (3), is true. We should note that the  
equations
\begin{equation}
\de v_r = \beta_3 (r\ep_r)^{1/3},
\label{rsh3}
\end{equation}
\begin{equation}
\de u_r = \beta_4 (r{{\Omega}}_r)^{1/3}
\label{rsh4}
\end{equation}
are both dimensionally plausible, and would yield
\begin{equation}
\zeta_p^{L} = \frac{p}{3} + o_{p/3},
\label{rsh3-e}
\end {equation}
\begin{equation}
\zeta_p^{T} = \frac{p}{3} + \tau_{p/3}.
\label{rsh4-e}
\end {equation}
Equation (4) would derive greater support if the alternative  
relations (7) and (8) can be shown to be unsatisfactory. 

\bigskip
\psfig{file=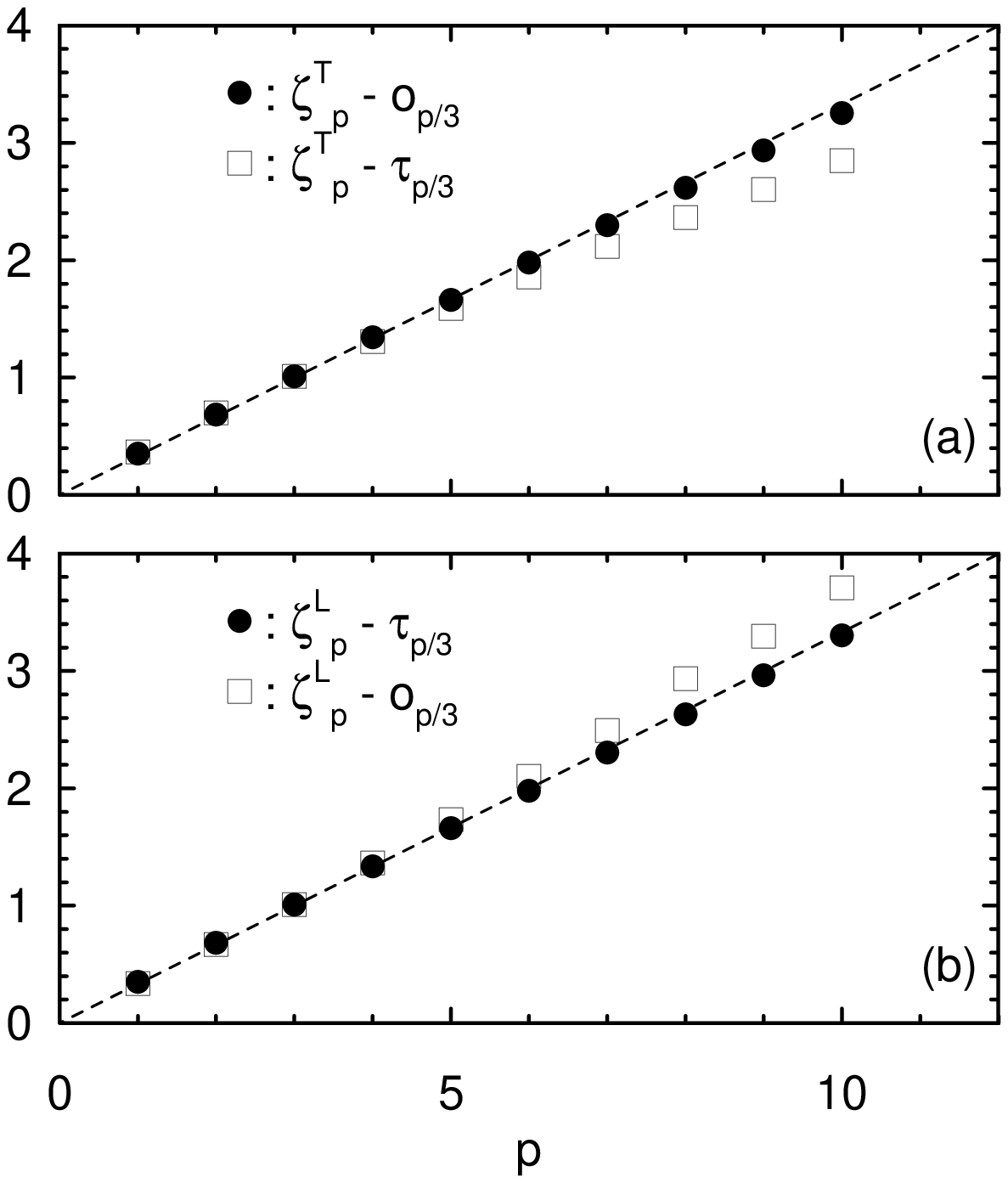,width=220pt}
\noindent
{\small FIG.~4. Verification of the scaling relations in RSHT. (a)
shows the results for the transverse structure functions and (b)
for the longitudinal case. The dashed lines correspond to $p/3$, which
would be followed by the numerical data if the similarity hypothesis
used is correct. See the equation pairs (2) and (4), and (7) and
(8).}
\bigskip

In Figs.~4 (a) and (b), we show comparisons of numerical results for  
$p$ up to 10 with predictions of (\ref{rsh2-e}) and (\ref{rsh4-e})  
for the transverse case, and with predictions of (\ref{rsh1-e}) and  
(\ref{rsh3-e}) for the longitudinal case. (For odd order, we have  
used absolute values of the appropriate velocity differences.) The  
exponents $\tau_{p/3}$ and $o_{p/3}$ used in the above relations were  
taken from previously published results for the same data  
\cite{cao,vor}. For $p < 3$, the differences between the implied  
relations (\ref{rsh1-e}), (\ref{rsh2-e}), and
(\ref{rsh3-e}), (\ref{rsh4-e}) and 
numerical results are relatively small, and cannot be  
distinguished easily. On the other hand, with increasing $p$, only  
$\zeta_p^{L} - \tau_{p/3}$ and $\zeta_p^{T} - o_{p/3}$ coincide with  
$p/3$, showing that Eqs. (\ref{rsh1-e}) and (\ref{rsh2-e}) are valid  
to a good approximation. The numerical evidence that $\zeta_p^{L} -  
o_{p/3}$ and $\zeta_p^{T} - \tau_{p/3}$ depart from $p/3$ for large  
$p$ effectively disqualifies (\ref{rsh3}) and (\ref{rsh4}). Note that  
the error bars in Figs. 4 are quite small and do not affect this  
conclusion. 

A more detailed test of Eq. (\ref{rsh2}) is given in Fig.~5, where  
we plot the normalized probability density function (PDF) of $\beta_2  
\equiv \Delta v_r/(r{{\Omega}}_r)^{1/3}$ with $r = 0.2$, $0.34$ and 
$0.44$, all of which lie in the inertial range. 
The three PDFs collapse quite well. It  
is seen that in the region $\beta_2/\la\beta_2^2\ra^{1/2} <3$, the  
PDFs agree with the standard Gaussian distribution. For  
larger values of $\beta_2/\la (\beta_2)^2 \ra^{1/2}$, the PDFs seem  
to depart slightly from Gaussian, with a tendency perhaps towards  
exponential. This trend has also been seen for the longitudinal  
counterpart, $\beta_1$, investigated by Wang {\em et al.}  
\cite{chen1}. One difference between the PDFs of $\beta_1$ and  
$\beta_2$, which follows from Kolmogorov's 4/5-ths law \cite{kolm},  
is that $\beta_2$ is not expected to be skewed, whereas $\beta_1$  
should have a third moment equal to $-4/5$. The PDFs of $\beta_3$ and  
$\beta_4$ have  
also been obtained but not shown here. They do not collapse for  
inertial range separations; nor do they agree with the Gaussian  
distribution even in the core region; and they possess larger  
departures from Gaussian near the tails.

The results in Figs.~4 and Fig.~5 suggest that RSHT proposed in  
(\ref{rsh2}) is a good working approximation connecting the  
transverse velocity increment with the enstrophy field. In addition,  
the differences of the intermittency properties between $\de u_r$ and  
$\de v_r$ as 
well as $\ep_r$ and ${\Omega}_r$ strongly suggest the possibility of  
two independent scaling groups. Note that $\ep_r$ and  
${\Omega}_r$ are the symmetric and antisymmetric parts of the  
strain-rate tensor, respectively; so we expect that this  
difference, reflecting the difference in the inertial-range physics  
of the dissipation and vortex dynamics, may quite plausibly carry  
over to the high Reynolds number limit as well. 

It should be pointed out that a more general refined similarity  
hypothesis, which encompasses results for both velocity components,  
cannot be ruled out. One possible scenario could be
\begin{equation}
\Delta v_r = \beta_5(r {{\Omega}}_r^{\alpha} \ep_r^{1 -  
\alpha})^{1/3}, 
\label{gene}
\end{equation}
where $\beta_5$ is again a universal stochastic variable and
$\alpha$ could, in general, be a function of the 
order index $p$\cite{titi}. Equation (\ref{gene}) is more  
complex than (\ref{rsh1}) and (\ref{rsh2}), since the scalings of the  
cross-correlation functions $\la {\Omega}_r^p \epsilon_r^q \ra$ have  
to be taken into account. 

\bigskip
\psfig{file=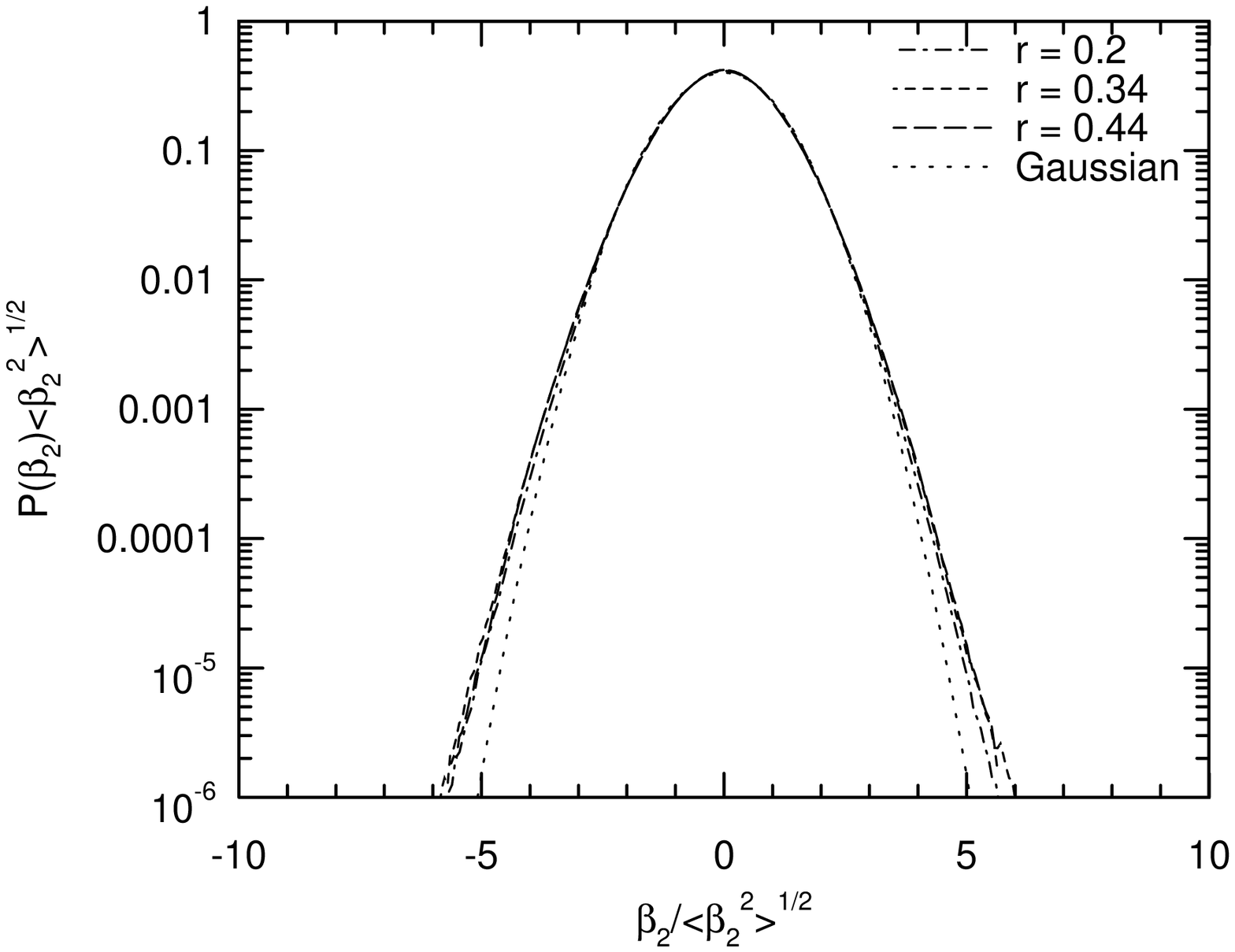,width=220pt}
\noindent
{\small FIG.~5.
Normalized PDFs of $\beta_2/\la(\beta_2)^2\ra^{1/2}$ for $r= 0.2$,
$0.34$ and $0.44$.}
\bigskip

In summary, we have proposed a refined similarity hypothesis for the  
transverse direction (RSHT); it connects the statistics of the  
transverse structure function in the inertial range with the locally  
averaged enstrophy. Data analysis of DNS for the Navier-Stokes  
equations at moderate Reynolds numbers demonstrate that RSHT is valid  
for the transverse structure functions in the inertial range, while  
Kolmogorov's RSH has been shown previously to be essentially correct  
for the longitudinal case \cite{sreeni1,pras,chen1}; the latter  
result is confirmed again in this paper. The important implication of RSHT is  
the possibility of the existence of two independent scaling groups  
which may correspond to different intermittent physics in fluid  
turbulence: these are related to the symmetric part of the strain-rate, or the  
dissipation physics, and those related to the antisymmetric part of  
the strain-rate, or the vortex dynamics. This view also suggests that 
no more than two sets of independent exponents are required for 
describing the scaling of all small-scale features as longitudinal 
and transverse velocity increments, dissipation, enstrophy, 
and circulation \cite{cao2}.

We thank G. D. Doolen, E. Titi, Y. Tu and Y. Zhou for useful  
discussions. Numerical simulations were carried out at the Advanced  
Computing Laboratory at Los Alamos National Laboratory and the Army  
High Performance Computing Research Center, using the Connection  
Machine-5. A grant from the National Science Foundation supported the  
research work of KRS.

%
%
%
%
%
%

\end{multicols}


\begin{references}
\bibitem{ess1} In applying ESS, we have plotted various orders of  
transverse structure functions against the third-order longitudinal  
structure function. Hyperviscous calculations also reveal differences  
between longitudinal and transverse exponents, but it would be  
difficult to use the hyperviscous data for modifying RSH.
\end{references}
\end{document}